\documentclass[aps,prb,twocolumn,superscriptaddress,showpacs,floatfix]{revtex4-1}

\usepackage[latin9]{inputenc}
\usepackage{verbatim}
\usepackage{amsmath}
\usepackage{amssymb}
\usepackage{graphicx}
\usepackage{esint}
\usepackage{hyperref}

\def\nn{\nonumber}

\def\beq{\begin{eqnarray}}
\def\eeq{\end{eqnarray}}

\def\up{\uparrow}
\def\down{\downarrow}

\let\baraccent=\= 
\renewcommand{\=}[1]{\stackrel{#1}{=}} 

\makeatletter

\pdfpageheight\paperheight
\pdfpagewidth\paperwidth

\@ifundefined{textcolor}{}
{%
 \definecolor{BLACK}{gray}{0}
 \definecolor{WHITE}{gray}{1}
 \definecolor{RED}{rgb}{1,0,0}
 \definecolor{GREEN}{rgb}{0,1,0}
 \definecolor{BLUE}{rgb}{0,0,1}
 \definecolor{CYAN}{cmyk}{1,0,0,0}
 \definecolor{MAGENTA}{cmyk}{0,1,0,0}
 \definecolor{YELLOW}{cmyk}{0,0,1,0}
}

\makeatother

\begin{document}

\title{Electrical detection of topological quantum phase transitions in disordered Majorana nanowires}

\author{Benjamin M. Fregoso}

\affiliation{Department of Physics, University of California, Berkeley, Berkeley, California 94720, USA.}

\affiliation{Condensed Matter Theory Center and Joint Quantum Institute, Department of Physics, University of Maryland, College Park, Maryland 20742-4111, USA.}

\author{Alejandro M. Lobos}

\affiliation{Condensed Matter Theory Center and Joint Quantum Institute, Department of Physics, University of Maryland, College Park, Maryland 20742-4111, USA.}

\author{S. Das Sarma}

\affiliation{Condensed Matter Theory Center and Joint Quantum Institute, Department of Physics, University of Maryland, College Park, Maryland 20742-4111, USA.}

\date{\today}
\begin{abstract}
We study a disordered superconducting nanowire, with broken time-reversal
and spin-rotational symmetry, which can be driven into a topological
phase with end Majorana bound states by an externally applied magnetic
field. It is known that as a function of disorder strength the Majorana
nanowire has a delocalization quantum phase transition from a topologically
non-trivial phase, which supports Majorana bound states, to a non-topological
insulating phase without them. On both sides of the transition, the
system is localized at zero energy albeit with very different topological
properties. We propose an electrical transport measurement to
detect the localization-delocalization transition occurring in the
bulk of the nanowire. The basic idea consists of measuring the \emph{difference}
of conductance at one end of the wire obtained at different values
of the coupling to the \emph{opposite} lead. We show that this measurement
reveals the non-local correlations emergent only at the topological transition. Hence,
while the proposed experiment does not directly probe the end Majorana
bound states, it can provide direct evidence for the bulk topological
quantum phase transition itself. 
\end{abstract}

\pacs{73.63.Nm, 74.45.+c, 74.81.-g, 03.65.Vf}

\maketitle
\textit{Introduction.} The study of topological phases of matter is
one of the most active research topics in all of physics~\cite{Nayak08}.
A recent proposal to realize a one-dimensional (1D) topological superconductor (SC)~\cite{kitaev2001}
supporting zero-energy Majorana bound states (MBSs) in semiconductor-superconductor
heterostructures \cite{Lutchyn2010,Oreg2010} has attracted great
deal of attention, and has been explored experimentally \cite{Mourik12_Signatures_of_MF,Das12_Evidence_of_MFs,Deng12_ZBP_in_Majorana_NW,Rokhinson2012,Finck13_ZBP_in_hybrid_NW_SC_device,Churchill2013}.
However, despite this excitement, the issue remains largely open and the need of more decisive (i.e., ``smoking gun'') evidence for
the MBS scenario has been emphasized in recent works \cite{Pikulin12_ZBP_from_weak_antilocalization_in_Majorana_NW,Bagrets12_Class_D_spectral_peak_in_Majorana_NW,Kells12_Near-zero-end_states_in_Majorana_wires_with_smooth_confinement,Rieder12_Endstates_in_multichannel_p-wave_SC_nanowires,Motrunich01_Disorder_in_topological_1D_SC,Liu12_ZBP_in_Majorana_wires_with_and_without_MZBSs,DasSarma12_Majorana_smoking_gun,Sau2013,Appelbaum2013}. In particular, no direct evidence for a topological quantum phase
transition (TQPT), which is characterized by the closing of the
superconducting gap and should accompany the emergence of
MBSs, has been detected so far.

Disorder (e.g., impurities in the semiconductor) is an important relevant
perturbation in the Majorana experiments \cite{Pikulin12_ZBP_from_weak_antilocalization_in_Majorana_NW,Bagrets12_Class_D_spectral_peak_in_Majorana_NW,Liu12_ZBP_in_Majorana_wires_with_and_without_MZBSs,Sau2013,Lobos2012,Rainis2013}, since the system is effectively a spinless p-wave superconductor with
no Anderson theorem\cite{Gennes1966}. Disorder and localization effects in superconductors
with broken time-reversal and spin-rotational symmetries [i.e., symmetry
class $D$ (Ref. \onlinecite{Altland1997})] have been a subject of intense theoretical
study \cite{Motrunich01_Disorder_in_topological_1D_SC,Brouwer00_Localization_Dirty_SC_wire,Gruzberg05_Localization_in_disordered_SC_wires_with_broken_SU2_symmetry,Brouwer11_Probability_distribution_of_MFS_in_disordered_wires,Brouwer11_Topological_SC_in_disorder_wires}.
In the work by Motrunich \textit{et al.}, it was shown that disorder-induced subgap Andreev
bound states proliferate in a class $D$ superconductor near zero
energy, and therefore a closing of the bulk SC gap at the TQPT
is an ill-defined concept since gap closing has no meaning 
if the system is already gapless.\cite{Motrunich01_Disorder_in_topological_1D_SC}.
Nevertheless, the system still has well-defined topological properties and generically
lies in one of two topologically distinct phases.
For weak disorder, an infinite system is in a non-trivial topological
phase characterized by the presence of two degenerate zero-energy
MBSs localized at the ends of the wire. In a finite-length system of
size $L$, this degeneracy is lifted by an exponential splitting $\sim e^{-L/\xi}$,
where $\xi$ is the superconducting coherence length. Increasing the
strength of disorder induces a proliferation of low-energy Andreev
bound states (i.e., quantum Griffiths effect), 
and the splitting scales as $\sim e^{L/\xi + L/(2\ell_e)}$, where $\ell_e$ is
the elastic mean-free path of the system.\cite{Brouwer11_Probability_distribution_of_MFS_in_disordered_wires} 
Beyond a critical
disorder strength (defined by the condition $\ell_e = \xi/2$), the 
system enters a nontopological insulating phase with no end-MBSs.
At both sides of the TQPT, the system is localized at zero energy,
and exactly at the critical point separating these phases, the wave
functions become delocalized and the smallest Lyapunov exponent (i.e.,
the inverse of the localization length of the system) vanishes. %
This key observation links the physics of localization and topological
properties of a disordered $D$-class SC wire \cite{Akhmerov11_Quantized_conductance_in_disordered_wire,DeGottardi11_MFs_with_disorder,DeGottardi_MFs_with_spatially_varying_potentials,Sau12_Robust_Majorana_chain,Adagideli13_Topological_order_in_dirty_wires}.
For example, it has been shown that the quantity $Q=\text{sign }\left(\text{Det }r\right)=\text{sign}(\prod_{n=1}^{2M}\tanh\lambda_{n})$,
where $r$ is the reflection matrix and $\lambda_{n}$ is the Lyapunov
exponent related to the $n$-th transmission channel, is a suitable
topological invariant for a disordered class $D$ superconductor,
which changes sign whenever a Lyapunov exponent crosses zero \cite{Akhmerov11_Quantized_conductance_in_disordered_wire,Fulga11_Scattering_formula_for_tological_wires}.
It was suggested that the delocalized nature of the wave functions,
with their non-local correlations that appear at the topological critical
point, could be observed in the quantization of the thermal conductance
$G_{\text{th}}/G_{0}=1$ (with $G_{0}=\pi^{2}k_{B}^{2}T/6h$ at temperature
$T$) or the onset of quantized non-local current-noise correlations, constituting
evidence for the TQPT. Unfortunately, the highly challenging requirements
of these experiments have hindered further progress.

\begin{figure}[t]
\includegraphics[scale=0.9]{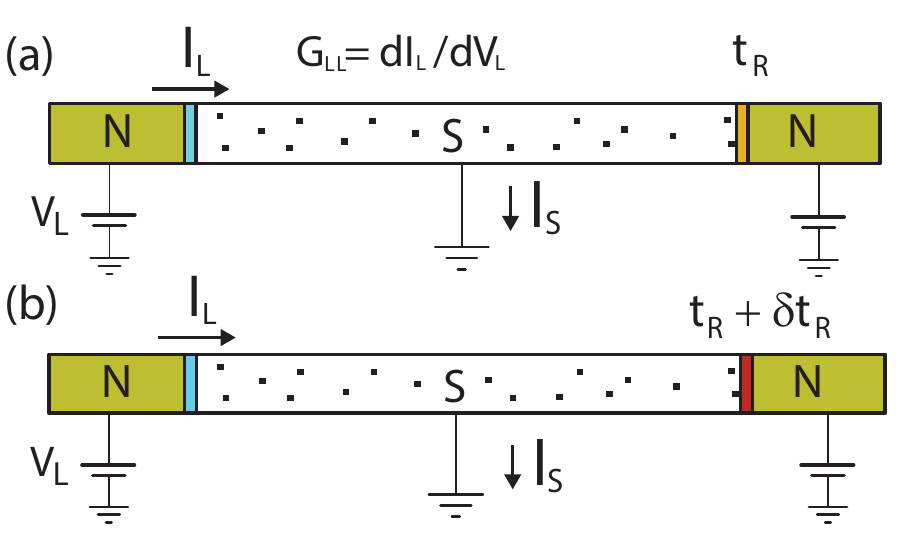} \caption{\label{fig:system} (Color online) Schematic representation of the N-S-N system under
consideration. The basic idea consists of measuring the conductance
at one end of the NW (e.g. left end) for different values of $\gamma_{R}$,
the coupling to the opposite (i.e., right) lead. The difference of
conductances $\delta G_{LL}$ [see Eq.~(\ref{eq:deltaGLL})] contains
information about the non-local correlations between the end-points
of the NW, which  can be used to detect the TQPT.}
\end{figure}

In this work we propose a different yet simple, electrical transport experiment
to detect the TQPT, measuring directly the non-local correlations
in the bulk appearing exactly at the critical point. We study a disordered
topological SC coupled to left and right normal leads in a normal-superconducting-normal (NSN) device
as depicted in Fig.~\ref{fig:system}. Instead of computing the left-right
conductance $G_{LR}$, our proposal consists in calculating the local
conductance \textit{at one end} of the NW, while tuning the coupling
to the \textit{opposite} lead. As we show below, this procedure allows
to extract information about the non-local correlations, which in
turn could be used to identify the TQPT in the bulk of the wire. We
stress that this is different from measuring $G_{LR}$ , which vanishes,
or from measuring the non-local correlations in the shot noise \cite{Bolech2007}.
Our method can be immediately implemented in on-going experiments
looking for zero bias conductance peak in the Majorana nanowires. 

\textit{Theoretical model.}
We firstmotivate our results by studying a 1D 
solvable model of a disordered $D$-class SC 
consisting of spinless  Dirac fermions with a random $p$-wave gap $\Delta(x)$.
In the Majorana basis the Hamiltonian, from $x=0$  to $x=L$, is
\cite{Brouwer11_Probability_distribution_of_MFS_in_disordered_wires,Akhmerov11_Quantized_conductance_in_disordered_wire} 
$H_{\text{w}}=-i\hbar v_{F}\sigma^{z}\partial_{x}+\sigma^{y}\Delta(x)$, where
$(\sigma^{x},\sigma^{y},\sigma^{z})$ is the vector of Pauli matrices acting on the space of right- and left-moving Majorana fields.
At zero energy, this Hamiltonian has localized Majorana modes at the
ends of the wire, e.g.,
$\Psi(x)=\exp[-(1/\hbar v_{F})\int_{0}^{x}dx^{\prime}\ \Delta\left(x^{\prime}\right)\sigma^{x}]\Psi\left(0\right)$ 
is a localized mode at the left end (i.e., $x=0$).
The reflection $r_{1}=\tanh(L\bar{\Delta}/\hbar v_{F})$ and transmission  
$t_{1}=\cosh^{-1}(L\bar{\Delta}/\hbar v_{F})$ amplitudes
are obtained
by imposing  $\Psi(0)=(1,r_1)^{T}$, $\Psi(L)=(t_{1},0)^{T}$ where
$\bar{\Delta}=L^{-1}\int_{0}^{L}dx\ \Delta\left(x\right)$ is the average $p$-wave gap.
Assuming that $\bar{\Delta}$ can be controlled
with an external tuning parameter (e.g., external magnetic field),
the TQPT in this model occurs when $\bar{\Delta}=0$, and is accompanied
by a change of sign in $r_1$ (which can be interpreted as
the topological invariant), and by a peak in the thermal conductance
$G_{\text{th}}/G_{0}=\text{Tr }t_1t_1^{*}=\cosh^{-2}(L\bar{\Delta}/\hbar v_{F})$,
of width equal to the Thouless energy of the system, i.e., $\hbar v_{F}/L$.
This result is a consequence of the particular reflection-less
boundary condition imposed at the right end of the wire, $x=L$. However,
one can assume a more general situation introducing a barrier at the
end of the wire, described by a generic scatterer with reflection
and transmission amplitudes $r_{2}$ and $t_{2}$, respectively (subject
to the unitarity constraint $\left|r_{2}\right|^{2}+\left|t_{2}\right|^{2}=1$).
This would correspond to an imperfect coupling
to the right lead or to any backscatterer which is \textit{external}
to the wire itself. The total reflection amplitude becomes 
\begin{align}
r & =r_{1}+t_{1}\left(\frac{r_{2}}{1+r_{2}r_{1}}\right)t_{1}.\label{eq:r_generic}
\end{align}
Intuitively, the last term in Eq.~(\ref{eq:r_generic}) represents processes
in which the right-moving Majorana mode 
is transmitted to the right end of the wire with amplitude $t_1$, and is 
reflected back with amplitude $r_2$ as a left-mover. 
This result means that the quantity $r$ in general contains
non-local contributions from the scattering occurring at the right
end\cite{Buttiker86_Landauer_Buttiker_paper}. Note that for a non-vanishing
$r_{2}$ the point $r=0$ is shifted with respect to the topological
transition in the bulk of the wire $\bar{\Delta}=0$. 
This shift is of the order of  the intrinsic width $\sim\hbar v_{F}/L$ of the TQPT, and hence does not affect its experimental detection.
Let us now assume that $r_{2}$ is another external tunable parameter
in the system, in which case a small variation $\delta r_{2}$ around
$r_{2}=0$ allows to extract the non-local contributions in Eq.~(\ref{eq:r_generic}),
i.e., $\delta r\approx\delta r_{2}\cosh^{-2}(L\bar{\Delta}/\hbar v_{F})$,
which is non-vanishing only near the TQPT, indicating the delocalization
of the Majorana wave function $\Psi\left(x\right)$.
Since these properties depend only on the
symmetry class of the Hamiltonian, we expect these findings
to be model independent and to apply to \textit{all types} class-$D$ Bogoliubov-de Gennes (BdG) Hamiltonians.
This is the main idea of this work. 

We now consider a more realistic model for a $D-$class superconducting
wire consisting of a 1D semiconductor NW of length $L$ along the $x$
axis with a strong spin-orbit coupling (SOC), an external magnetic
field along $x$, and proximity induced $s-$wave pairing due to a
proximate bulk SC \cite{Lutchyn2010,Oreg2010}. Discretizing the continuum
system and assuming single subband occupancy, the effective low-energy
model corresponds to an $N-$site tight-binding model 
$H=H_{\text{w}}+H_{\text{leads}}+H_{\text{mix}}$~\cite{Stanescu11_MFs_in_SM_nanowires},
with

\begin{align}
H_{\text{w}}= & -t\sum_{\langle lm\rangle,s}c_{l,s}^{\dagger}c_{m,s}-\sum_{l,s}c_{l,s}^{\dagger}\left(\mu_{l}-V_{Z}\sigma_{ss^{\prime}}^{x}\right)c_{l,s^{\prime}}\nonumber \\
 & +\sum_{l,s}\left(i\alpha\ c_{l,s}^{\dagger}\sigma_{ss^{\prime}}^{y}c_{l+1,s^{\prime}}+\Delta_{0}c_{l\uparrow}^{\dagger}c_{l\downarrow}^{\dagger}+{\rm H.c.}\right),\label{eq:Hw}
\end{align}
with effective hopping parameter $t$ and lattice parameter $a$.
Here $c_{l,s}^{\dagger}$ creates an electron
with spin projection $s=\left\{ \uparrow,\downarrow\right\} $ at
site $l$ in the tight-binding chain, $\alpha$ is the Rashba SOC
parameter, $V_{Z}$ is the Zeeman energy due to an external magnetic
field along $x$, and $\Delta_{0}$ is the induced $s$-wave gap which
must be calculated self-consistently. Since the precise numerical value of
$\Delta_{0}$ does not modify our conclusions,
here we make the simplifying assumption that $\Delta_{0}$ already satisfies the self-consistent
SC gap equation.

Short-ranged nonmagnetic static disorder in the semiconductor NW
is included through a fluctuating chemical potential $\mu_{l}=\mu_{0}+\delta\mu_{l}$
about the average $\mu_{0}$. For simplicity we assume $\delta\mu_{l}$
to be a delta-correlated random variable with Gaussian distribution,
$\left\langle \delta\mu_{l}\delta\mu_{m}\right\rangle =v_{0}^{2}\delta_{lm}$.
Hamiltonian in Eq.~\ref{eq:Hw} is another particular example of a disordered
$D-$class SC \cite{Altland1997}. As a function of the external Zeeman
field $V_{Z}$, and in absence of disorder this model has a TQPT from
a topologically trivial phase to non-trivial phase with end MBS at
the value $V_{Z,c}=\sqrt{\mu_{0}^{2}+\Delta_{0}^{2}}$ as shown originally
by Sau \textit{et al.} \cite{Sau2010}. In the presence of disorder, the critical
field $V_{Z,c}$ shifts to higher values, and its precise value depends
on the particular details of the disorder \cite{Brouwer11_Probability_distribution_of_MFS_in_disordered_wires,Brouwer11_Topological_SC_in_disorder_wires,Adagideli13_Topological_order_in_dirty_wires}.

We describe the coupling to the external leads (see Fig.~\ref{fig:system}),
by the term $H_{\text{mix}}=\sum_{s}(t_{L}f_{Lk,s}^{\dagger}c_{1,s}+t_{R}f_{Rk,s}^{\dagger}c_{N,s})+\text{H.c.,}$
where $t_{L\left(R\right)}$ is the coupling to the left (right) lead
and $f_{L\left(R\right)k,s}^{\dagger}$ is the corresponding creation
operator for fermions with quantum number $k$ and spin $s$. The
external leads are modeled as large Fermi liquids with Hamiltonian
$H_{\text{lead},j}=\sum_{k,s}\epsilon_{k}f_{j,k,s}^{\dagger}f_{j,k,s}$,
where $j=\{L,R\}$. 

At $T=0$, the local and non-local zero-bias conductances 
have the explicit form \cite{Blonder1982,Anantram96_Andreev_scattering} (see Appendix~\ref{app:sns})
\begin{eqnarray}
G_{LL}&=&\frac{e^{2}}{h}\{ M_{L}-\text{Tr }[r_{ee}r_{ee}^{\dagger}]+\text{Tr }[r_{eh}r_{eh}^{\dagger}]\},\label{eq:GLL}\\
G_{LR}&=&-\frac{e^{2}}{h}\{ \text{Tr }[t_{ee}t_{ee}^{\dagger}]-\text{Tr }[t_{eh}t_{eh}^{\dagger}]\},\label{eq:GLR}
\end{eqnarray}
where $M_{L}=\sum_{\sigma}2\pi\gamma_{L}\rho_{1\sigma}(0)$
is the number of transmission channels at energy $\omega=0$ in the
left lead. Here $\rho_{l\sigma}\left(\omega\right)$ is the local
density of states at site $l$ in the chain, and $\gamma_{j}\equiv2\pi t_{j}^{2}\rho_{j}^{0}$
is the broadening of levels due to the leakage to the lead $j$, described
by the local density of states $\rho_{j}^{0}$ (assumed to be $SU(2)$-symmetric
and constant around the Fermi energy). In addition, we have defined,
respectively, the normal and Andreev reflection matrices at the left
lead, i.e., $\left[r_{ee}\left(\omega\right)\right]_{s,s^{\prime}}\equiv\gamma_{L}g_{1s,1s^{\prime}}^{r}\left(\omega\right)$,
$\left[r_{eh}\left(\omega\right)\right]_{s,s^{\prime}}\equiv\gamma_{L}f_{1s,1s^{\prime}}^{r}\left(\omega\right)$,
and the normal and Andreev transmission matrices, i.e., $\left[t_{ee}\left(\omega\right)\right]_{s,s^{\prime}}\equiv\sqrt{\gamma_{L}\gamma_{R}}g_{1s,Ns^{\prime}}^{r}\left(\omega\right)$
and $\left[t_{eh}\left(\omega\right)\right]_{s,s^{\prime}}\equiv\sqrt{\gamma_{L}\gamma_{R}}f_{1s,Ns^{\prime}}^{r}\left(\omega\right)$,
where $g_{ls,ms^{\prime}}^{r}\left(\omega\right)$ and $f_{ls,ms^{\prime}}^{r}\left(\omega\right)$
are the normal and anomalous retarded Green's functions in the chain (see Appendix~\ref{app:sns}).

The topological phase occurring for $V_{Z}>V_{Z,c}$ is
characterized by a quantized zero-bias peak at $G_{LL}=2e^{2}/h$,
which is a direct consequence of an MBS localized at the left end
of the NW \cite{Law09,Sau_long,Flensberg10_Quantization_MBS,Wimmer}.
However, the proliferation of disorder-induced subgap Andreev bound
states near zero energy results in a power-law singularity $\left\langle \rho_{1,\sigma}\left(\omega\right)\right\rangle _{\text{dis}}\sim1/\left|\omega\right|^{\nu}$
in the disorder-averaged density of states, and complicates the interpretation
of this zero-bias peak \cite{Motrunich01_Disorder_in_topological_1D_SC,Sau2013,Brouwer00_Localization_Dirty_SC_wire,Gruzberg05_Localization_in_disordered_SC_wires_with_broken_SU2_symmetry,Brouwer11_Probability_distribution_of_MFS_in_disordered_wires,Brouwer11_Topological_SC_in_disorder_wires}.
On the other hand, the experimental detection of the predicted delocalization
TQPT is hindered by the fact that the non-local electrical conductance
$G_{LR}\left(0\right)$ vanishes, since $\text{Tr }[t_{ee}t_{ee}^{\dagger}]=\text{Tr }[t_{eh}t_{eh}^{\dagger}]$
at the transition \cite{Akhmerov11_Quantized_conductance_in_disordered_wire}. In addition, the predicted quantized thermal conductance
$G_{\text{th}}/G_{0}=2\text{Tr }[t_{ee}t_{ee}^{\dagger}+t_{eh}t_{eh}^{\dagger}]$
is experimentally very difficult to observe. This calls for alternative methods to detect the TQPT. 

In analogy to Eq.~(\ref{eq:r_generic}), Eq.~(\ref{eq:GLL}), despite
being a local quantity computed at the left lead, contains information
about the non-local correlations in the NW. To see this, we make use of the Green's function identity\cite{Kadanoff1989}
$\boldsymbol{\mathcal{G}}^{r}\left(\omega\right)-\boldsymbol{\mathcal{G}}^{a}\left(\omega\right)=\boldsymbol{\mathcal{G}}^{r}\left(\omega\right)\left[\boldsymbol{\Sigma}^{r}\left(\omega\right)-\boldsymbol{\Sigma}^{a}\left(\omega\right)\right]\boldsymbol{\mathcal{G}}^{a}\left(\omega\right)$,
where $\boldsymbol{\mathcal{G}}^{r\left(a\right)}\left(\omega\right)=[\omega-\boldsymbol{\mathcal{H}}_{\text{w}}^{\text{BdG}}-\boldsymbol{\Sigma}^{r\left(a\right)}]^{-1}$
is the Green's function matrix, defined in terms of the $2\times2$
Nambu blocks 
\beq
\boldsymbol{\mathcal{G}}^{r\left(a\right)}\left(\omega\right)=\left(\begin{array}{cc}
g^{r}\left(\omega\right) & f^{r}\left(\omega\right)\\
\bar{f}^{r}\left(\omega\right) & \bar{g}^{r}\left(\omega\right)
\end{array}\right), \nn
\eeq 
$\boldsymbol{\mathcal{H}}_{\text{w}}^{\text{BdG}}$ is the BdG Hamiltonian
corresponding to Eq.~(\ref{eq:Hw}), and $\boldsymbol{\Sigma}^{r\left(a\right)}\left(\omega\right)=\mp(i/2)\left(\gamma_{L}\delta_{l,1}+\gamma_{R}\delta_{l,N}\right)\delta_{s,s^{\prime}}$
is the retarded (advanced) self-energy due to the coupling $H_{\text{mix}}$.
This allows us to express Eq.~(\ref{eq:GLL}) in a more suggestive
form
\begin{eqnarray}
G_{LL} & =\frac{e^{2}}{h}\{2\text{Tr }[r_{eh}r_{eh}^{\dagger}]+\text{Tr }[t_{ee}t_{ee}^{\dagger}+t_{eh}t_{eh}^{\dagger}]\},\label{eq:GLL2}
\end{eqnarray}
which is reminiscent to Eq.~(\ref{eq:r_generic}), and where last
term is the (dimensionless) thermal conductance $G_{\text{th}}/G_{0}$. This term vanishes 
in the limit $L~\to~\infty$, where we recover the usual expression $G_{LL}=(2e^2/h) \text{Tr}[r_{eh}r_{eh}^{\dagger}]$ found in the literature \cite{Law09, Wimmer}.
Changing the coupling to the right lead, $\gamma_{R} \to \gamma_{R}+\delta\gamma_{R}$
(keeping all the other parameters fixed) amounts to varying the reflection amplitude
$r_{2}$ in the continuum model, and hence we expect to obtain non-local correlations at the TQPT. Experimentally, $\gamma_{R}$ and
$\gamma_{L}$ could be easily modified varying the pinch-off gates underneath
the ends of the NW, constituting a useful experimental knob in the Majorana experiment, 
which has not been exploited in Refs. \cite{Mourik12_Signatures_of_MF,Das12_Evidence_of_MFs,Deng12_ZBP_in_Majorana_NW,Rokhinson2012,Finck13_ZBP_in_hybrid_NW_SC_device,Churchill2013}.
In particular, one can easily show that the change in $G_{LL}$ at zero bias,
\begin{align}
\delta G_{LL} & \equiv G_{LL}\left(\gamma_{R}+\delta\gamma_{R},\gamma_{L}\right)-G_{LL}\left(\gamma_{R},\gamma_{L}\right),\label{eq:deltaGLL}
\end{align}
is a purely non-local contribution proportional to $\boldsymbol{\mathcal{G}}_{1\sigma,N\sigma^{\prime}}^{r\left(a\right)}\left(\omega\right)$,
$\boldsymbol{\mathcal{G}}_{N\sigma,1\sigma^{\prime}}^{r\left(a\right)}\left(\omega\right)$
and $\boldsymbol{\mathcal{G}}_{N\sigma,N\sigma^{\prime}}^{r\left(a\right)}\left(\omega\right)$
(see appendix \ref{app:sns} for details). In agreement with our previous results, 
this contribution will be only non-vanishing when the single-particle wave functions become delocalized, allowing a simple electrical
detection of the TQPT. 

\begin{figure}
\includegraphics[scale=0.31]{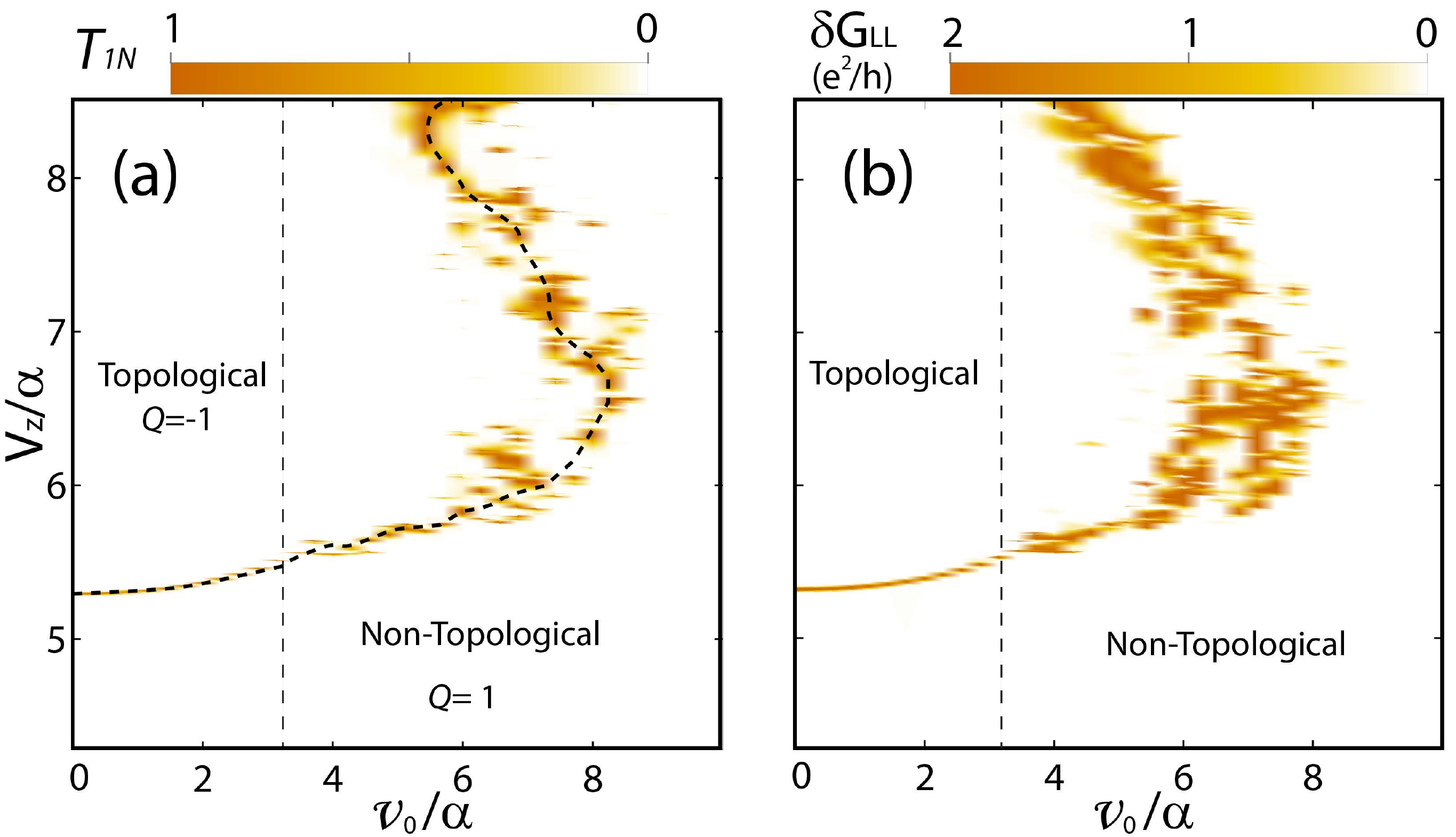} \caption{\label{fig:phase_diag}(Color online). (a) Transmission probability
$T_{1N}$ between the end-points (sites $1$ and $N$) of the isolated
NW and topological phase diagram as a function of disorder strength
$v_{0}$ and Zeeman field $V_{Z}$. At the topological critical point,
the topological invariant $Q=\text{sign}(\prod_{n=1}^{4}\tanh\lambda_{n})$,
where $\lambda_{n}$ are the Lyapunov exponents, changes sign and
the transmission probability $T_{1N}$ becomes 1 [see also Fig.~\ref{fig:deltaGLL2}(a)]. The bold 
dashed line follows the TQPT.
(b) Difference of left conductances $\delta G_{LL}$ [see Eq.~(\ref{eq:deltaGLL})].
We have chosen parameters $\gamma_{L}=0.2t$,
$\gamma_{R}=10^{-5}t$, $\delta\gamma_{R}=10^{-3}t$, $N=300$, $\alpha=0.07t$,
$\Delta=0.37t$ and $\mu_{0}=-2t$. In the clean case and at $V_{z}=0.45t=6.4\alpha$
we obtain the maximal ratio of $L/\xi\sim 10$ where $L$ is the length
of the NW.}
\end{figure}

\textit{Transfer matrix}. We consider a \textit{single}
disorder realization $\boldsymbol{\delta\mu}=\{\delta\mu_{1},\dots,\delta\mu_{N}\}$,
and vary an overall prefactor, the disorder strength $v_{0}$. 
Presumably, a fixed disorder realization is closer to the experiments,
where the semiconductor NW is in the mesoscopic regime, and it is
not clear that disorder necessarily self-averages at the very low
experimental temperatures. We have computed the topological phase diagram using 
the transfer matrix method\cite{DeGottardi11_MFs_with_disorder} for 
an isolated NW using the model Hamiltonian of Eq.~(\ref{eq:Hw}) 
The transfer matrix for zero-energy modes is given by $M=\prod_{l=1}^{N}M_{l}$,
where (see Appendix ~\ref{app:transfer})
\begin{eqnarray*}
M_{l} & = & \left(\begin{array}{cccc}
\frac{-t\mu_{l}+\alpha(V_{z}-\Delta_{0})}{t^{2}+\alpha^{2}} & \frac{-\alpha\mu_{l}+t(V_{z}+\Delta_{0})}{t^{2}+\alpha^{2}} & \frac{t}{t^{2}+\alpha^{2}} & \frac{\alpha}{t^{2}+\alpha^{2}}\\
\frac{\alpha\mu_{l}+t(V_{z}-\Delta_{0})}{t^{2}+\alpha^{2}} & \frac{-t\mu_{l}-\alpha(V_{z}+\Delta_{0})}{t^{2}+\alpha^{2}} & \frac{-\alpha}{t^{2}+\alpha^{2}} & \frac{t}{t^{2}+\alpha^{2}}\\
-t & -\alpha & 0 & 0\\
\alpha & -t & 0 & 0
\end{array}\right).
\end{eqnarray*}
The eigenvalues of $M$ are denoted by $e^{\pm\lambda_{n}}$, where
the Lyapunov exponent $\lambda_{n}$ is related to the transmission
probability by $T_{1N}=\sum_{n=1}^{4}T_{n}$, with $T_{n}=1/\cosh^{2}\lambda_{n}$
the transmission eigenvalue corresponding to the $n-$th channel.
At the TQPT one of the Lyapunov exponents in the NW vanishes and, consequently,
the corresponding transmission eigenvalue becomes $T_{n}=1$, while
the topological invariant $Q$ changes sign.

\textit{Disscussion}. In Fig.~\ref{fig:phase_diag}(a) we show the topological quantum phase
diagram in the Zeeman field vs disorder strength plane. 
Fig.~\ref{fig:phase_diag}(b) shows the $\delta G_{LL}$ conductance
map for exactly the same parameters as in Fig. \ref{fig:phase_diag}a.
Although $\delta G_{LL}$ is computed for an open system,
while $T_{1N}$ has been computed for the isolated wire ($\gamma_{L}=\gamma_{R}=0$),
the remarkable agreement between Fig.~\ref{fig:phase_diag}(a) and
\ref{fig:phase_diag}(b) is encouraging for the experimental detection
of the TQPT using $\delta G_{LL}$.

\begin{figure}
\includegraphics[scale=0.44]{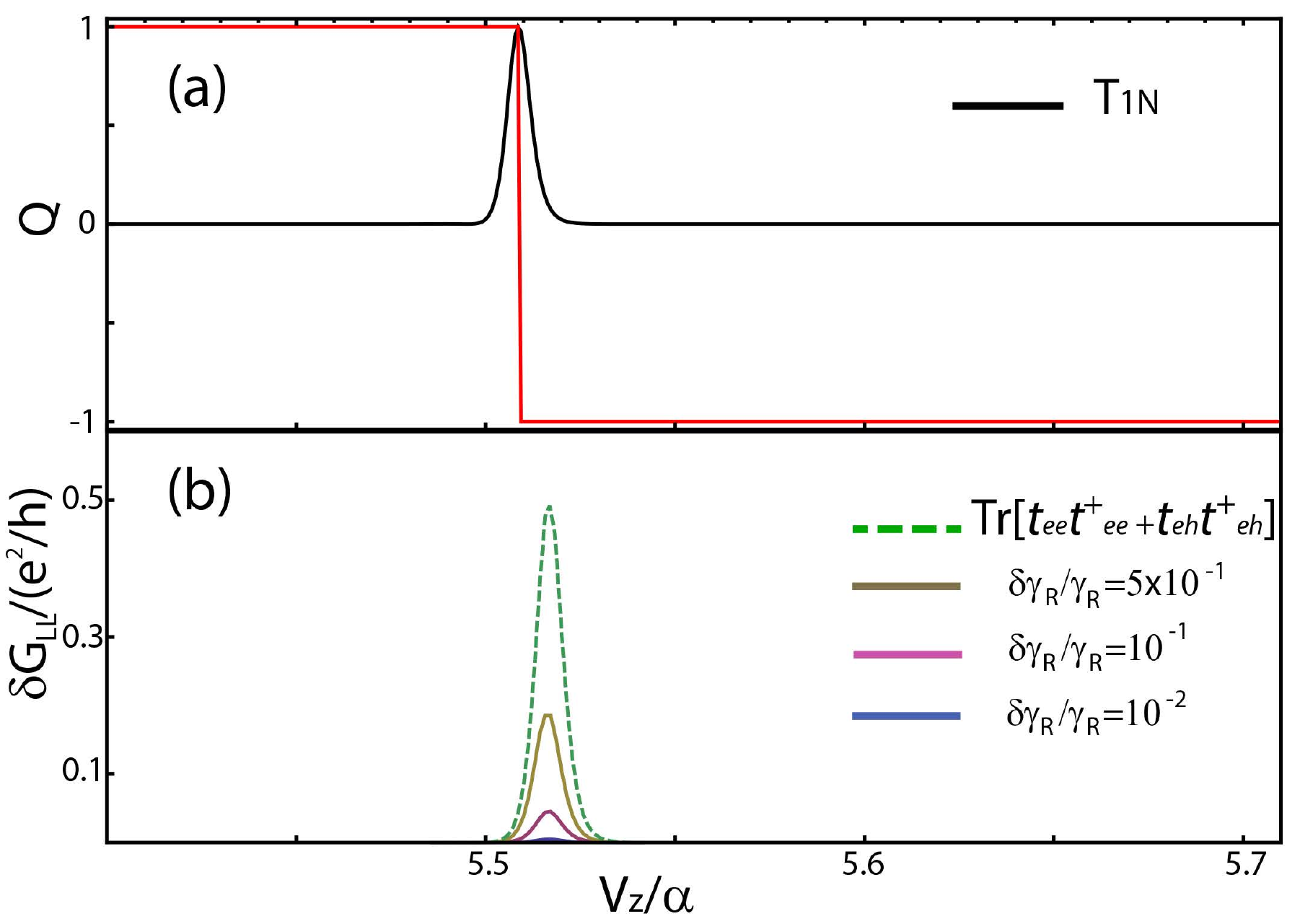} \caption{\label{fig:deltaGLL2}(Color online). (a) Topological invariant for
an isolated NW. The localization -delocalization transition is evident
form the appearance of a quantized peak in the transmission probability,
$T_{1N}$. (b) difference in conductance in the left lead for different
couplings to the right lead in the regime $\delta\gamma_{R}\ll\gamma_{R}$.
The main point is that these two quantities follow each other which
can be used to detect the topological phase transition. Here we have
chosen fixed disorder amplitude $v_{0}=3.14\alpha$, see dashed vertical lines
in Fig.~\ref{fig:phase_diag}.}
\end{figure}

In Fig.~\ref{fig:deltaGLL2}(a) and \ref{fig:deltaGLL2}(b) we compare the transmission probability
$T_{1N}$ and the topological invariant $Q$ for the isolated NW for a particular 
disorder strength with 
$\delta G_{LL}$ in the limit
$\delta\gamma_{R}\ll\gamma_{R}=\gamma_{L}=1.4\alpha$, for various
$\delta\gamma_{R}$. As we see, these two quantities follow each other closely. 
While the width of the peaks is the same in all cases (as expected, since the Thouless energy $\hbar v_F/L$ is an intrinsic property of $H_\text{w}$), the
maximum of $\delta G_{LL}$ is shifted with respect to the maximum
of $T_{1N}$, indicative of some reflection occurring at the NS barriers.

In practice, our proposal is expected to work best for short
wires, where the maximal ratio $L/\xi$ is not too large.
Note that the visibility of the electrical signal crucially depends on the width
$~\hbar v_{F}/L$ of the peak in $\delta G_{LL}$. A very narrow peak might be
hard to detect, or could be washed away by finite temperature effects or other
dissipative mechanisms not taken into consideration here. Also,
the system should be smaller than the phase-relaxation length $L< L_{\phi}$.
Despite these limitations, our predictions are 
within experimental reach\cite{Mourik12_Signatures_of_MF,Das12_Evidence_of_MFs,Deng12_ZBP_in_Majorana_NW,Rokhinson2012,Finck13_ZBP_in_hybrid_NW_SC_device,Churchill2013}
since we obtain $\xi \approx 20$ nm with the experimental value of $L\approx 2$ $\mu$m. 
Similarly, the width of the peak in $\delta G_{LL}$, proportional to the Thouless energy, is
of the order of $\hbar v_F/L \approx 52 $ $\mu$eV, also within experimental
resolution.

In conclusion, we have developed a method to experimentally detect
the topological phase transition in disordered class $D$ SC NWs,
like those under investigation in Refs. \cite{Mourik12_Signatures_of_MF,Das12_Evidence_of_MFs,Deng12_ZBP_in_Majorana_NW,Rokhinson2012,Finck13_ZBP_in_hybrid_NW_SC_device,Churchill2013}.
While this method cannot provide direct evidence for MBS, it is can
provide robust evidence of the topological phase transition itself
in disordered NWs. The basic idea is to measure the differences of
conductance at one end of the NW (e.g., the left end) for different
values of the coupling with the opposite lead.  We note that this procedure can be easily implemented in on-going
experiments and provides a complementary technique of studying topological
physics in Majorana-carrying systems by directly studying the bulk
TQPT rather than the Majorana zero modes themselves. %

The authors thank L. Arrachea 
for useful comments and acknowledge support from DARPA QuEST, NSF through the PFC@JQI, Conacyt, and Microsoft Q.

\appendix
\onecolumngrid
\section{Calculation of the conductance matrix in a SNS contact}
\label{app:sns}
Here we provide the details of the calculations of the  conductance through a generic
NSN system. Our derivation is standard and makes use of the so-called Hamiltonian formalism 
\cite{Cuevas96_Hamiltonian_approach_to_SC_contacts}, which is equivalent to the the more 
frequently-used scattering or BTK formalism \cite{Blonder1982, Anantram96_Andreev_scattering},
provided the Green functions are calculated to all orders in the coupling across the SN interface. 
Our model Hamiltonian in the main text is

\begin{align}
H & =H_{\text{w}} +H_{\text{mix}}+H_\text{lead, L}+H_\text{lead, R},\label{eq:H}\\
H_{\text{w}} &= -t\sum_{\langle lm\rangle,s}c_{l,s}^{\dagger}c_{m,s}-\sum_{l,s}c_{l,s}^{\dagger}\left(\mu_{l}-V_{Z}\sigma_{ss^{\prime}}^{x}\right)c_{l,s^{\prime}}	 +\sum_{l,s}\left(i\alpha\ c_{l,s}^{\dagger}\sigma_{ss^{\prime}}^{y}c_{l+1,s^{\prime}}+\Delta_{0}c_{l\uparrow}^{\dagger}c_{l\downarrow}^{\dagger}+{\rm H.c.}\right), \\
H_{\text{mix}}&=\sum_{s}(t_{L}f_{Lk,s}^{\dagger}c_{1,s}+t_{R}f_{Rk,s}^{\dagger}c_{N,s})+\text{H.c.,}\label{eq:Hmix}\\
H_{\text{lead},j}&=\sum_{k,s}\epsilon_{k}f_{j,k,s}^{\dagger}f_{j,k,s}. \label{eq:Hlead}
\end{align}
We assume that each lead is in equilibrium at a chemical potential $\mu_{j}=eV_{j}$
controlled by external voltages, where $j=\left\{ L,R\right\} $,
and that the SC NW is grounded, i.e., $\mu_{S}=0$  (see Fig. 1). 
The expression for the electric current calculated through the contacts is  
$I_{j}=e\langle dN_{j}/dt\rangle =ie\langle [H,N_{j}]\rangle/\hbar =ie \langle [H_{\text{mix}},N_{j}]\rangle/\hbar $,
which can be written in terms of the  Green function at the
contacts as~\cite{Cuevas96_Hamiltonian_approach_to_SC_contacts,meir92}

\begin{align}
I_{L} & =\frac{ie}{\hbar}\sum_{\sigma}t_{L}\left[\left\langle c_{L,\sigma}^{\dagger}c_{1,\sigma}\right\rangle -\left\langle c_{1,\sigma}^{\dagger}c_{L,\sigma}\right\rangle \right],\label{eq:currentL}\\
I_{R} & =\frac{ie}{\hbar}\sum_{\sigma}t_{L}\left[\left\langle c_{R,\sigma}^{\dagger}c_{N,\sigma}\right\rangle -\left\langle c_{N,\sigma}^{\dagger}c_{R,\sigma}\right\rangle \right].\label{eq:currentR}
\end{align}
With these definitions, note that the currents are positive if particles move into the leads (i.e., exit the SC), and negative
otherwise. On the other hand, charge conservation demands that $I_{L}+I_{R}+I_{S}=0$,
where $I_{S}$ is the excess current that flows to earth through the
SC. Within the Baym-Kadanoff-Keldysh formalism \cite{Kadanoff1989} we define the lesser Green function

\begin{align}
g_{i\sigma,j\sigma^{\prime}}^{<}\left(t\right) & \equiv ie\left\langle c_{i,\sigma}^{\dagger}c_{j,\sigma}\left(t\right)\right\rangle ,
\end{align}
so that we can write the currents as
\begin{align}
I_{L}  & =\frac{e}{\hbar}t_{L}\sum_{\sigma}\int_{-\infty}^{\infty}\frac{d\omega}{2\pi}\left[g_{L\sigma,1\sigma}^{<}\left(\omega\right)-g_{1\sigma,L\sigma}^{<}\left(\omega\right)\right],\label{eq:currentL_2}\\
I_{R}  & =\frac{e}{\hbar}t_{R}\sum_{\sigma}\int_{-\infty}^{\infty}\frac{d\omega}{2\pi}\left[g_{R\sigma,N\sigma}^{<}\left(\omega\right)-g_{N\sigma,R\sigma}^{<}\left(\omega\right)\right],\label{eq:currentR_2}
\end{align}
Using equations of motion, we can express Eqs. \ref{eq:currentL_2}
and \ref{eq:currentR_2} in terms of local Green's functions as \cite{Cuevas96_Hamiltonian_approach_to_SC_contacts,meir92}

\begin{align}
I_{L} & =-\frac{e}{h}t_{L}^{2}\sum_{\sigma}\int_{-\infty}^{\infty}d\omega\left[g_{L\sigma,L\sigma}^{0,<}\left(\omega\right)g_{1\sigma,1\sigma}^{>}\left(\omega\right)-g_{L\sigma,L\sigma}^{0,>}\left(\omega\right)g_{1\sigma,1\sigma}^{<}\left(\omega\right)\right],\label{eq:currentL_3}\\
I_{R} & =-\frac{e}{h}t_{R}^{2}\sum_{\sigma}\int_{-\infty}^{\infty}d\omega\left[g_{R\sigma,R\sigma}^{0,<}\left(\omega\right)g_{N\sigma,N\sigma}^{>}\left(\omega\right)-g_{R\sigma,R\sigma}^{0,>}\left(\omega\right)g_{N\sigma,N\sigma}^{<}\left(\omega\right)\right].\label{eq:currentR_3}
\end{align}

Our first step to obtain the expression of the currents is to specify
the unperturbed Green's functions $g_{j\sigma,j\sigma}^{0,\gtrless}\left(\omega\right)$
in the leads, with $j=\left\{ L,R\right\} $:

\begin{align*}
g_{j\sigma,j\sigma}^{0,<}\left(\omega\right) & =2\pi i\rho_{j,\sigma}^{0}\left(\omega\right)n_{j}\left(\omega\right),\\
g_{j\sigma,j\sigma}^{0,>}\left(\omega\right) & =2\pi i\rho_{j,\sigma}^{0}\left(\omega\right)\left[n_{j}\left(\omega\right)-1\right],
\end{align*}
where $n_{j}\left(\omega\right)=n_{F}\left(\omega+\mu_{j}\right)$
are the Fermi distribution functions at the leads. Substituting these
expressions gives

\begin{align}
I_{L} & =-\frac{ie}{h}2\pi t_{L}^{2}\sum_{\sigma}\int_{-\infty}^{\infty}d\omega\;\rho_{L,\sigma}^{0}\left(\omega\right)\left\{ n_{L}\left(\omega\right)\left[g_{1\sigma,1\sigma}^{r}\left(\omega\right)-g_{1\sigma,1\sigma}^{a}\left(\omega\right)\right]+g_{1\sigma,1\sigma}^{<}\left(\omega\right)\right\} ,\label{eq:currentL_4}\\
I_{R} & =-\frac{ie}{h}2\pi t_{R}^{2}\sum_{\sigma}\int_{-\infty}^{\infty}d\omega\;\rho_{R,\sigma}^{0}\left(\omega\right)\left\{ n_{R}\left(\omega\right)\left[g_{N\sigma,N\sigma}^{r}\left(\omega\right)-g_{N\sigma,N\sigma}^{a}\left(\omega\right)\right]+g_{N\sigma,N\sigma}^{<}\left(\omega\right)\right\} ,\label{eq:currentR_4}
\end{align}
where we have used the identity\cite{Kadanoff1989,mahan} $g^{>}\left(\omega\right)-g^{<}\left(\omega\right)=g^{r}\left(\omega\right)-g^{a}\left(\omega\right)$.  
Obtaining an explicit expression for the currents $I_{L}$ and $I_{R}$
is quite cumbersome. Since we will be interested only in the conductance,
we note that there is an enormous simplification if we compute directly
the conductance matrix by deriving the currents with respect to the
voltages $V_{L},V_{R}$. Then
\begin{align}
G_{LL}\equiv\frac{dI_{L}}{dV_{L}} & =-\frac{ie^{2}}{h}2\pi t_{L}^{2}\sum_{\sigma}\int_{-\infty}^{\infty}d\omega\;\rho_{L,\sigma}^{0}\left(\omega\right)\left\{ \frac{dn_{L}\left(\omega\right)}{d\left(eV_{L}\right)}\left[g_{1\sigma,1\sigma}^{r}\left(\omega\right)-g_{1\sigma,1\sigma}^{a}\left(\omega\right)\right]+\frac{dg_{1\sigma,1\sigma}^{<}\left(\omega\right)}{d\left(eV_{L}\right)}\right\} ,\label{eq:G_LL}\\
G_{LR}\equiv\frac{dI_{L}}{dV_{R}} & =-\frac{ie^{2}}{h}2\pi t_{L}^{2}\sum_{\sigma}\int_{-\infty}^{\infty}d\omega\;\rho_{L,\sigma}^{0}\left(\omega\right)\frac{dg_{1\sigma,1\sigma}^{<}\left(\omega\right)}{d\left(eV_{R}\right)},\label{eq:G_LR}\\
G_{RL}\equiv\frac{dI_{R}}{dV_{L}} & =-\frac{ie^{2}}{h}2\pi t_{R}^{2}\sum_{\sigma}\int_{-\infty}^{\infty}d\omega\;\rho_{R,\sigma}^{0}\left(\omega\right)\frac{dg_{N\sigma,N\sigma}^{<}\left(\omega\right)}{d\left(eV_{L}\right)},\label{eq:G_RL}\\
G_{RR}\equiv\frac{dI_{R}}{dV_{R}} & =-\frac{ie^{2}}{h}2\pi t_{R}^{2}\sum_{\sigma}\int_{-\infty}^{\infty}d\omega\;\rho_{R,\sigma}^{0}\left(\omega\right)\left\{ \frac{dn_{R}\left(\omega\right)}{d\left(eV_{R}\right)}\left[g_{N\sigma,N\sigma}^{r}\left(\omega\right)-g_{N\sigma,N\sigma}^{a}\left(\omega\right)\right]+\frac{dg_{N\sigma,N\sigma}^{<}\left(\omega\right)}{d\left(eV_{R}\right)}\right\} ,\label{eq:G_RR}
\end{align}

Therefore, we see that the problem is reduced to finding the 
Green's functions in the superconducting system. In a non-interacting system, the full Green's function verifies the Dyson's equation in Nambu space \cite{Cuevas96_Hamiltonian_approach_to_SC_contacts}

\begin{align}
\boldsymbol{\mathcal{G}}^{\gtrless}\left(\omega\right) & =\left[\mathbf{1}+\boldsymbol{\mathcal{G}}^{r}\left(\omega\right)\left(\boldsymbol{\mathcal{T}}_{L}+\boldsymbol{\mathcal{T}}_{R}\right)\right]\boldsymbol{\mathcal{G}}^{0,\gtrless}\left(\omega\right)\left[\mathbf{1}+\left(\boldsymbol{\mathcal{T}}_{L}+\boldsymbol{\mathcal{T}}_{R}\right)\boldsymbol{\mathcal{G}}^{a}\left(\omega\right)\right],\label{eq:G_Dyson_Keldysh}\\
\boldsymbol{\mathcal{G}}^{\left(r,a\right)}\left(\omega\right) & =\boldsymbol{\mathcal{G}}^{0,\left(r,a\right)}\left(\omega\right)+\boldsymbol{\mathcal{G}}^{0,\left(r,a\right)}\left(\omega\right)\left(\boldsymbol{\mathcal{T}}_{L}+\boldsymbol{\mathcal{T}}_{R}\right)\boldsymbol{\mathcal{G}}^{\left(r,a\right)}\left(\omega\right),\label{eq:G_Dyson_ra}
\end{align}
where we have introduced the Nambu notation
\begin{align}
\boldsymbol{\mathcal{G}}_{i\sigma,j\sigma^{\prime}}^{\nu}\left(z\right) & =\left(\begin{array}{cc}
g_{i\sigma,j\sigma^{\prime}}^{\nu}\left(z\right) & f_{i\sigma,j\sigma^{\prime}}^{\nu}\left(z\right)\\
\bar{f}_{i\sigma,j\sigma^{\prime}}^{\nu}\left(z\right) & \bar{g}_{i\sigma,j\sigma^{\prime}}^{\nu}\left(z\right)
\end{array}\right),
\end{align}
with $\nu=\left\{ >,<,r,a\right\} $, and where
\begin{align}
\boldsymbol{\mathcal{T}} & _{j}=\left(\begin{array}{cc}
t_{j} & 0\\
0 & -t_{j}
\end{array}\right).
\end{align}
The unperturbed Green's functions (i.e., computed for $t_{L}=t_{R}=0$)
are 

\begin{align}
\boldsymbol{\mathcal{G}}_{i\sigma,j\sigma^{\prime}}^{0,<}\left(\omega\right) & =2\pi i\boldsymbol{\rho}_{i\sigma,j\sigma^{\prime}}^{0}\left(\omega\right)n_{F}\left(\omega\right),\label{eq:G0_lesser}\\
\boldsymbol{\mathcal{G}}_{i\sigma,j\sigma^{\prime}}^{0,>}\left(\omega\right) & =2\pi i\boldsymbol{\rho}_{i\sigma,j\sigma^{\prime}}^{0}\left(\omega\right)\left[n_{F}\left(\omega\right)-1\right],\label{eq:G0_bigger}\\
\boldsymbol{\rho}_{i\sigma,j\sigma^{\prime}}^{0}\left(\omega\right) & =-\frac{1}{\pi}\text{Im}\left[\boldsymbol{\mathcal{G}}_{i\sigma,j\sigma^{\prime}}^{0,r}\left(\omega\right)\right]=\left(\begin{array}{cc}
\rho_{i\sigma,j\sigma^{\prime}}^{0}\left(\omega\right) & \zeta_{i\sigma,j\sigma^{\prime}}^{0}\left(\omega\right)\\
\zeta_{i\sigma,j\sigma^{\prime}}^{0}\left(\omega\right) & \bar{\rho}_{i\sigma,j\sigma^{\prime}}^{0}\left(\omega\right)
\end{array}\right),\label{eq:Rho}
\end{align}
We only need the derivative with respect to the voltages, which are
only in the leads. This gives,

\begin{align*}
\frac{dg_{1\sigma,1\sigma}^{\gtrless}}{d\left(eV_{L}\right)} & =2\pi it_{L}^{2}\sum_{s}\left[\frac{dn_{L}}{d\left(eV_{L}\right)}\rho_{L}^{0}g_{1\sigma,1s}^{r}g_{1s,1\sigma}^{a}+\frac{d\bar{n}_{L}}{d\left(eV_{L}\right)}\bar{\rho}_{L}^{0}f_{1\sigma,1s}^{r}\bar{f}_{1s,1\sigma}^{a}\right],\\
\frac{dg_{1\sigma,1\sigma}^{\gtrless}}{d\left(eV_{R}\right)} & =2\pi it_{R}^{2}\sum_{s}\left[\frac{dn_{R}}{d\left(eV_{R}\right)}\rho_{R}^{0}g_{1\sigma,Ns}^{r}g_{Ns,1\sigma}^{a}+\frac{d\bar{n}_{R}}{d\left(eV_{R}\right)}\bar{\rho}_{R}^{0}f_{1\sigma,Ns}^{r}\bar{f}_{Ns,1\sigma}^{a}\right],\\
\frac{dg_{N\sigma,N\sigma}^{\gtrless}}{d\left(eV_{L}\right)} & =2\pi it_{L}^{2}\sum_{s}\left[\frac{dn_{L}}{d\left(eV_{L}\right)}\rho_{L}^{0}g_{N\sigma,1s}^{r}g_{1s,N\sigma}^{a}+\frac{d\bar{n}_{L}}{d\left(eV_{L}\right)}\bar{\rho}_{L}^{0}f_{N\sigma,1s}^{r}\bar{f}_{1s,N\sigma}^{a}\right],\\
\frac{dg_{N\sigma,N\sigma}^{\gtrless}}{d\left(eV_{R}\right)} & =2\pi it_{R}^{2}\sum_{s}\left[\frac{dn_{R}}{d\left(eV_{R}\right)}\rho_{R}^{0}g_{N\sigma,Ns}^{r}g_{Ns,N\sigma}^{a}+\frac{d\bar{n}_{R}}{d\left(eV_{R}\right)}\bar{\rho}_{R}^{0}f_{N\sigma,Ns}^{r}\bar{f}_{Ns,N\sigma}^{a}\right].
\end{align*}
Substituting into Eqs. \ref{eq:G_LL}-\ref{eq:G_RR},
and using the result $g_{j\sigma,j\sigma}^{r}\left(\omega\right)-g_{j\sigma,j\sigma}^{a}\left(\omega\right)=-2\pi i\rho_{j\sigma}\left(\omega\right)$, where we have defined the local density of states $\rho_{j\sigma}\left(\omega\right)\equiv \rho_{j\sigma,j\sigma}\left(\omega\right)$, yields

\begin{eqnarray}
G_{LL} & =&-\frac{e^{2}}{h}\sum_{\sigma}\int_{-\infty}^{\infty}d\omega\;\gamma_{L}\left(\omega\right)\left[\frac{dn_{L}}{d\left(eV_{L}\right)}2\pi\rho_{1\sigma}-\sum_{s}\frac{dn_{L}}{d\left(eV_{L}\right)}\gamma_{L}g_{1\sigma,1s}^{r}g_{1s,1\sigma}^{a}-\sum_{s}\frac{d\bar{n}_{L}}{d\left(eV_{L}\right)}\bar{\gamma}_{L}f_{1\sigma,1s}^{r}\bar{f}_{1s,1\sigma}^{a}\right]_{\omega},\label{eq:GLL_2}\\
G_{LR} & =&\frac{e^{2}}{h}\sum_{\sigma,s}\int_{-\infty}^{\infty}d\omega\;\left[\gamma_{L}\gamma_{R}\frac{dn_{R}}{d\left(eV_{R}\right)}g_{1\sigma,Ns}^{r}g_{Ns,1\sigma}^{a}+\frac{d\bar{n}_{R}}{d\left(eV_{R}\right)}\gamma_{L}\bar{\gamma}_{R}f_{1\sigma,Ns}^{r}\bar{f}_{Ns,1\sigma}^{a}\right]_{\omega},\label{eq:GLR_2}\\
G_{RL} & =&\frac{e^{2}}{h}\sum_{\sigma,s}\int_{-\infty}^{\infty}d\omega\;\left[\frac{dn_{L}\left(\omega\right)}{d\left(eV_{L}\right)}\gamma_{R}\gamma_{L}g_{N\sigma,1s}^{r}g_{1s,N\sigma}^{a}+\frac{d\bar{n}_{L}}{d\left(eV_{L}\right)}\gamma_{R}\bar{\gamma}_{L}f_{N\sigma,1s}^{r}\bar{f}_{1s,N\sigma}^{a}\right]_{\omega},\label{eq:GRL_2}\\
G_{RR} & =&-\frac{e^{2}}{h}\sum_{\sigma}\int_{-\infty}^{\infty}d\omega\;\gamma_{R}\left(\omega\right)\bigg[\frac{dn_{R}}{d\left(eV_{R}\right)}2\pi\rho_{N\sigma}-\sum_{s}\frac{dn_{R}}{d\left(eV_{R}\right)}\gamma_{R}g_{N\sigma,Ns}^{r}g_{Ns,N\sigma}^{a}\nn \\
&&\hspace{250pt}-\sum_{s}\frac{d\bar{n}_{R}}{d\left(eV_{R}\right)}\bar{\gamma}_{R}f_{N\sigma,Ns}^{r}\bar{f}_{Ns,N\sigma}^{a}\bigg]_{\omega},\label{eq:GRR_2}
\end{eqnarray}
where we have defined the broadening
\begin{align}	
\gamma_j\left(\omega\right) & =2\pi t_j^{2}\rho_j^{0}\left(\omega\right),\\
\bar{\gamma}_j\left(\omega\right) & =2\pi t_j^{2}\bar{\rho}_j^{0}\left(\omega\right).
\end{align}
In particular at $T=0$ and zero-bias, and assuming electron-hole symmetry in the leads (i.e., $\gamma_j=\bar{\gamma}_j$), we obtain

\begin{align}
G_{LL} & =\frac{e^{2}}{h}\sum_{\sigma}\left[2\pi\gamma_{L}\rho_{1\sigma}-\sum_{s}\gamma_{L}^{2}\left|g_{1\sigma,1s}^{r}\right|^{2}+\sum_{s}\gamma_{L}^{2}\left|f_{1\sigma,1s}^{r}\right|^{2}\right]_{\omega=0},\label{eq:GLL_final}\\
G_{LR} & =-\frac{e^{2}}{h}\sum_{\sigma,s}\gamma_{L}\gamma_{R}\left[\left|g_{1\sigma,Ns}^{r}\right|^{2}-\left|f_{1\sigma,Ns}^{r}\right|^{2}\right]_{\omega=0},\label{eq:GLR_final}\\
G_{RL} & =-\frac{e^{2}}{h}\sum_{\sigma,s}\gamma_{R}\gamma_{L}\left[\left|g_{N\sigma,1s}^{r}\right|^{2}-\left|f_{N\sigma,1s}^{r}\right|^{2}\right]_{\omega=0},\label{eq:GRL_final}\\
G_{RR} & =\frac{e^{2}}{h}\sum_{\sigma}\left[2\pi\gamma_{R}\rho_{N\sigma}-\sum_{s}\gamma_{R}^{2}\left|g_{N\sigma,Ns}^{r}\right|^{2}+\sum_{s}\gamma_{R}^{2}\left|f_{N\sigma,Ns}^{r}\right|^{2}\right]_{\omega=0},\label{eq:GRR_final}
\end{align}

To make contact with BTK theory \cite{Blonder1982, Anantram96_Andreev_scattering}, we can express these results in a more standard 
form by recalling that $M_{L}=2\pi \text{Tr }\left[\boldsymbol{\Gamma}_{L}\boldsymbol{\rho}_{1}\right]=\sum_{\sigma}2\pi\gamma_{L}\rho_{1\sigma}\left(\omega\right)$
is the number of modes in the lead $L$, and $M_{R}=2\pi \text{Tr }\left[\boldsymbol{\Gamma}_{R}\boldsymbol{\rho}_{N}\right]=\sum_{\sigma}2\pi\gamma_{R}\rho_{N\sigma}\left(\omega\right)$,
where we have defined the matrices $\boldsymbol{\Gamma}_{L(R)}=\left(\begin{array}{cc}
\gamma_{L(R)} & 0\\
0 & \gamma_{L(R)}
\end{array}\right)$, and $\boldsymbol{\rho}_{1(N)}=2\pi\left(\begin{array}{cc}
\rho_{L(N),\uparrow}\left(\omega\right) & 0\\
0 & \rho_{L(N),\downarrow}\left(\omega\right)
\end{array}\right)$(see Ref.~\onlinecite{datta}). On the other hand, defining the matrices
\[
\begin{array}{cclcccl}
\mathbf{r}_{ee}^{LL} & =&\left(\begin{array}{cc}
\gamma_{L}g_{1\uparrow,1\uparrow}^{r} & \gamma_{L}g_{1\uparrow,1\downarrow}^{r}\\
\gamma_{L}g_{1\downarrow,1\uparrow}^{r} & \gamma_{L}g_{1\downarrow,1\downarrow}^{r}
\end{array}\right)_{\omega} & \qquad & \mathbf{r}_{eh}^{LL}&= & \left(\begin{array}{cc}
\gamma_{L}f_{1\uparrow,1\uparrow}^{r} & \gamma_{L}f_{1\uparrow,1\downarrow}^{r}\\
\gamma_{L}f_{1\downarrow,1\uparrow}^{r} & \gamma_{L}f_{1\downarrow,1\downarrow}^{r}
\end{array}\right)_{\omega}\\
\mathbf{r}_{ee}^{RR} & =&\left(\begin{array}{cc}
\gamma_{R}g_{N\uparrow,N\uparrow}^{r} & \gamma_{R}g_{N\uparrow,N\downarrow}^{r}\\
\gamma_{R}g_{N\downarrow,N\uparrow}^{r} & \gamma_{R}g_{N\downarrow,N\downarrow}^{r}
\end{array}\right)_{\omega} & \qquad & \mathbf{r}_{eh}^{RR}&= & \left(\begin{array}{cc}
\gamma_{R}f_{N\uparrow,N\uparrow}^{r} & \gamma_{R}f_{N\uparrow,N\downarrow}^{r}\\
\gamma_{R}f_{N\downarrow,N\uparrow}^{r} & \gamma_{R}f_{N\downarrow,N\downarrow}^{r}
\end{array}\right)_{\omega}\\
\mathbf{t}_{ee}^{LR} & =&\left(\begin{array}{cc}
\sqrt{\gamma_{L}\gamma_{R}}g_{1\uparrow,N\uparrow}^{r} & \sqrt{\gamma_{L}\gamma_{R}}g_{1\uparrow,N\downarrow}^{r}\\
\sqrt{\gamma_{L}\gamma_{R}}g_{N\downarrow,N\uparrow}^{r} & \sqrt{\gamma_{L}\gamma_{R}}g_{1\downarrow,N\downarrow}^{r}
\end{array}\right)_{\omega} & \qquad & \mathbf{t}_{eh}^{LR}&= & \left(\begin{array}{cc}
\sqrt{\gamma_{L}\gamma_{R}}f_{1\uparrow,N\uparrow}^{r} & \sqrt{\gamma_{L}\gamma_{R}}f_{1\uparrow,N\downarrow}^{r}\\
\sqrt{\gamma_{L}\gamma_{R}}f_{N\downarrow,N\uparrow}^{r} & \sqrt{\gamma_{L}\gamma_{R}}f_{1\downarrow,N\downarrow}^{r}
\end{array}\right)_{\omega}
\end{array}\]
we can express our Eqs. \ref{eq:GLL_final}-\ref{eq:GRR_final}
in the BTK language as \cite{Blonder1982, Anantram96_Andreev_scattering}

\begin{align}
G_{LL} & =\frac{e^{2}}{h}\left\{ M_{L}-\text{Tr }\left[\mathbf{r}_{ee}^{LL}\left(\mathbf{r}_{ee}^{LL}\right)^{\dagger}\right]+\text{Tr }\left[\mathbf{r}_{eh}^{LL}\left(\mathbf{r}_{eh}^{LL}\right)^{\dagger}\right]\right\} _{\omega=0},\label{eq:GLL_BTK}\\
G_{LR} & =-\frac{e^{2}}{h}\left\{ \text{Tr }\left[\mathbf{t}_{ee}^{LR}\left(\mathbf{t}_{ee}^{LR}\right)^{\dagger}\right]-\text{Tr }\left[\mathbf{t}_{eh}^{LR}\left(\mathbf{t}_{eh}^{LR}\right)^{\dagger}\right]\right\} _{\omega=0},\label{eq:GLR_BTK}\\
G_{RL} & =-\frac{e^{2}}{h}\left\{ \text{Tr }\left[\mathbf{t}_{ee}^{RL}\left(\mathbf{t}_{ee}^{RL}\right)^{\dagger}\right]-\text{Tr }\left[\mathbf{t}_{eh}^{RL}\left(\mathbf{t}_{eh}^{RL}\right)^{\dagger}\right]\right\} _{\omega=0},\label{eq:GRL_BTK}\\
G_{RR} & =\frac{e^{2}}{h}\left\{ M_{R}-\text{Tr }\left[\mathbf{r}_{ee}^{RR}\left(\mathbf{r}_{ee}^{RR}\right)^{\dagger}\right]+\text{Tr }\left[\mathbf{r}_{eh}^{RR}\left(\mathbf{r}_{eh}^{RR}\right)^{\dagger}\right]\right\} _{\omega=0},\label{eq:GRR_BTK}
\end{align}

In order to make explicit the non-local terms in these expressions we make use of the identity \cite{datta}
\begin{align}
\boldsymbol{\mathcal{G}}^{r}\left(\omega\right)-\boldsymbol{\mathcal{G}}^{a}\left(\omega\right) & =\boldsymbol{\mathcal{G}}^{r}\left(\omega\right)\left[\boldsymbol{\Sigma}^{r}\left(\omega\right)-\boldsymbol{\Sigma}^{a}\left(\omega\right)\right]\boldsymbol{\mathcal{G}}^{a}\left(\omega\right),
\end{align}
From here, the following results are obtained
\begin{align}
g_{1\sigma,1\sigma}^{r}-g_{1\sigma,1\sigma}^{a} & =-2\pi i\rho_{1,\sigma}\\
&=-2\pi i\sum_{s}\left[t_{L}^{2}\rho_{L}^{0}g_{1\sigma,1s}^{r}g_{1s,1\sigma}^{a}+t_{L}^{2}\bar{\rho}_{L}^{0}f_{1\sigma,1s}^{r}\bar{f}_{1s,1\sigma}^{a}+t_{R}^{2}\rho_{R}^{0}g_{1\sigma,Ns}^{r}g_{Ns,1\sigma}^{a}+t_{R}^{2}\bar{\rho}_{R}^{0}f_{1\sigma,Ns}^{r}\bar{f}_{Ns,1\sigma}^{a}\right],\\
g_{N\sigma,N\sigma}^{r}-g_{N\sigma,N\sigma}^{a} & =-2\pi i\rho_{N,\sigma}\\
&=-2\pi i\sum_{s}\left[t_{R}^{2}\rho_{R}^{0}g_{N\sigma,Ns}^{r}g_{Ns,N\sigma}^{a}+t_{R}^{2}\bar{\rho}_{R}^{0}f_{N\sigma,Ns}^{r}\bar{f}_{Ns,N\sigma}^{a}+t_{L}^{2}\rho_{L}^{0}g_{N\sigma,1s}^{r}g_{1s,N\sigma}^{a}+t_{L}^{2}\bar{\rho}_{L}^{0}f_{N\sigma,1s}^{r}\bar{f}_{1s,N\sigma}^{a}\right],
\end{align}
and hence, substituting into Eqs. \ref{eq:GLL_2}-\ref{eq:GRR_2},
we obtain 

\begin{align}
G_{LL} & =\frac{e^{2}}{h}\sum_{\sigma,s}\left[2\gamma_{L}^{2}f_{1\sigma,1s}^{r}\bar{f}_{1s,1\sigma}^{a}+\gamma_{L}\gamma_{R}\left(g_{1\sigma,Ns}^{r}g_{Ns,1\sigma}^{a}+f_{1\sigma,Ns}^{r}\bar{f}_{Ns,1\sigma}^{a}\right)\right],\label{eq:GLL_final_symm}\\
G_{LR} & =-\frac{e^{2}}{h}\sum_{\sigma,s}\gamma_{L}\gamma_{R}\left[g_{1\sigma,Ns}^{r}g_{Ns,1\sigma}^{a}-f_{1\sigma,Ns}^{r}\bar{f}_{Ns,1\sigma}^{a}\right],\label{eq:GLR_final_symm}\\
G_{RL} & =-\frac{e^{2}}{h}\sum_{\sigma,s}\gamma_{R}\gamma_{L}\left[g_{N\sigma,1s}^{r}g_{1s,N\sigma}^{a}-f_{N\sigma,1s}^{r}\bar{f}_{1s,N\sigma}^{a}\right],\label{eq:GRL_final_symm}\\
G_{RR} & =\frac{e^{2}}{h}\sum_{\sigma,s}\left[2\gamma_{R}^{2}f_{N\sigma,Ns}^{r}\bar{f}_{Ns,N\sigma}^{a}+\gamma_{R}\gamma_{L}\left(g_{N\sigma,1s}^{r}g_{1s,N\sigma}^{a}+f_{N\sigma,1s}^{r}\bar{f}_{1s,N\sigma}^{a}\right)\right],\label{eq:GRR_final_symm}
\end{align}
In particular, Eq.~\ref{eq:GLL_final_symm} corresponds to Eq.~5 in the main text.

\section{Transmission probability and topological phase diagram for a closed system obtained via the Transfer Matrix method}
\label{app:transfer}
The equations of motion for the fermionic operators 
$c_{n,\sigma}, c^\dagger_{n,\sigma}$ in the isolated $N$-site NW (see Eq.~\ref{eq:Hw}) are

\beq
i\frac{d}{dt} c_{n,\up} &=& -t (c_{n+1,\up} +c_{n-1,\up}) -\mu_n c_{n,\up} + V_{z,n} c_{n,\down} + \alpha (c_{n+1,\down} -c_{n-1,\down}) + \Delta_n c_{n,\down}^{\dagger} \\
i\frac{d}{dt} c_{n,\down} &=& -t (c_{n+1,\down} +c_{n-1,\down}) -\mu_n c_{n,\down} + V_{z,n} c_{n,\up} - \alpha (c_{n+1,\up} -c_{n-1,\up}) - \Delta_n c_{n,\up}^{\dagger} 
\eeq
where we have included possible inhomogeneity in the chemical potential, paring potential 
and magnetic field (random hopping could also be easily incorporated).
In the Majorana basis $c_{n,\up} = (a_n + i b_n)/2, c_{n,\up}^{\dagger} = (a_n - i b_n)/2, c_{n,\down} = (\bar{a}_n + i \bar{b}_n)/2, 
c_{n,\down}^{\dagger} = (\bar{a}_n - i \bar{b}_n)/2$ the equations of motion are 

\beq
\frac{d}{dt} a_{n} &=& -t (b_{n+1} +b_{n-1}) -\mu_n b_{n} + V_{z,n} \bar{b}_{n} + \alpha (\bar{b}_{n+1} - \bar{b}_{n-1}) - \Delta_n \bar{b}_{n} \\
-\frac{d}{dt} b_{n} &=& -t (a_{n+1} +a_{n-1}) -\mu_n a_{n} + V_{z,n} \bar{a}_{n} + \alpha (\bar{a}_{n+1} - \bar{a}_{n-1}) + \Delta_n \bar{a}_{n} \\
\frac{d}{dt} \bar{a}_{n} &=& -t (\bar{b}_{n+1} +\bar{b}_{n-1}) -\mu_n \bar{b}_{n} + V_{z,n} b_{n} - \alpha (b_{n+1} - b_{n-1}) + \Delta_n b_{n} \\
-\frac{d}{dt} \bar{b}_{n} &=& -t (\bar{a}_{n+1} +\bar{a}_{n-1}) -\mu_n \bar{a}_{n} + V_{z,n} a_{n} - \alpha (a_{n+1} - a_{n-1}) - \Delta_n a_{n} 
\eeq
We are interested in the normal modes of the NW which we assume are linear combinations of Majorana operators,
$Q=\sum_{n} (\gamma_n a_n +  \bar{\gamma}_n \bar{a}_n +i \eta_n b_n +  i \bar{\eta}_n \bar{b}_n)$. For clarity  
we suppress a label indexing the modes. The coefficients $\gamma$'s and $\eta$'s 
are determined by requiring that the operator be an eigenmode of energy $E$, i.e., 
$i d Q/dt = E Q$. Using the fact the Majorana operators are complete and matching like 
terms we obtain the discrete form of the  Schrodinger equation

\beq
- E 
\begin{pmatrix}
 \eta_n  \\ 
\bar{\eta}_n 
\end{pmatrix}
&=&
\begin{pmatrix} -t & \alpha \\ -\alpha & -t \end{pmatrix} 
\begin{pmatrix} \gamma_{n+1} \\ \bar{\gamma}_{n+1} \end{pmatrix}  + 
\begin{pmatrix} -t & -\alpha \\ \alpha & -t \end{pmatrix} 
\begin{pmatrix} \gamma_{n-1} \\ \bar{\gamma}_{n-1} \end{pmatrix}  + 
\begin{pmatrix} -\mu_n & V_{z,n} + \Delta_n \\ V_{z,n} - \Delta_n & -\mu_n \end{pmatrix} 
\begin{pmatrix} \gamma_{n} \\ \bar{\gamma}_{n} \end{pmatrix},  \\
 - E 
\begin{pmatrix}
 \gamma_n  \\ 
\bar{\gamma}_n 
\end{pmatrix}
&=&
\begin{pmatrix} -t & \alpha \\ -\alpha & -t \end{pmatrix} 
\begin{pmatrix} \eta_{n+1} \\ \bar{\eta}_{n+1} \end{pmatrix}  + 
\begin{pmatrix} -t & -\alpha \\ \alpha & -t \end{pmatrix} 
\begin{pmatrix} \eta_{n-1} \\ \bar{\eta}_{n-1} \end{pmatrix}  + 
\begin{pmatrix} -\mu_n & V_{z,n} - \Delta_n \\ V_{z,n} + \Delta_n & -\mu_n \end{pmatrix} 
\begin{pmatrix} \eta_{n} \\ \bar{\eta}_{n} \end{pmatrix} ,
\eeq
Which is of the form

\beq
E\psi_n = \kappa_n \psi_{n+1} + \kappa^{\dagger}_{n-1}\psi_{n-1} + u_n \psi_{n}
\eeq
and can therefore be written as a \textit{transfer matrix}

\beq
\begin{pmatrix}
 \psi_{n+1}  \\ 
\kappa_n^{\dagger} \psi_n 
\end{pmatrix}.
&=&
\begin{pmatrix} -\kappa^{-1}_n (E-u_n) & -\kappa_n^{-1} \\ \kappa_n^{\dagger} & 0 \end{pmatrix} 
\begin{pmatrix} \psi_{n} \\ \kappa_{n-1}^{\dagger} \psi_{n-1} \end{pmatrix}   
\eeq 
At zero energy we can define two independent Majorana operators  
$Q_1 \equiv \sum_n (\gamma_n a_n + \bar{\gamma}_n \bar{a}_n)$ and 
$Q_2 \equiv \sum_n (\eta_n b_n + \bar{\eta}_n \bar{b}_n)$ which contain the same information about the localization properties of the  
system. Focusing on the transfer matrix for the $a_n, \bar{a}_n$ modes, we obtain  
at zero energy,
\beq
M_n &=& \begin{pmatrix} \kappa^{-1}_n u_n & -\kappa_n^{-1} \\ \kappa_n^{\dagger} & 0 \end{pmatrix} 
\label{tmatrix}
\\
\psi_{n}   &=& 
\begin{pmatrix} \gamma_{n} \\ \bar{\gamma}_n \end{pmatrix}  \\
\kappa_n &=& \begin{pmatrix} -t & \alpha \\ -\alpha & -t \end{pmatrix} \\
u_n &=& \begin{pmatrix}  -\mu_n & V_{z,n} + \Delta_n \\ V_{z,n} - \Delta_n & -\mu_n \end{pmatrix}
\eeq
and hence the $M_n$ matrix is a $4\times 4$ matrix,

\beq
M_n &=& 
\begin{pmatrix} 
\frac{-t \mu_n + \alpha (V_{z,n}-\Delta_n)}{t^2 + \alpha^2} & \frac{-\alpha \mu_n + t (V_{z,n}+\Delta_n)}{t^2 + \alpha^2} & \frac{t}{t^2 + \alpha^2} & \frac{\alpha}{t^2 + \alpha^2}  \\ 
\frac{\alpha \mu_n + t (V_{z,n}-\Delta_n)}{t^2 + \alpha^2} & \frac{-t \mu_n - \alpha (V_{z,n}+\Delta_n)}{t^2 + \alpha^2} & \frac{-\alpha}{t^2 + \alpha^2} & \frac{t}{t^2 + \alpha^2}  \\ 
-t & -\alpha & 0 & 0 \\
\alpha & -t & 0 & 0 
\end{pmatrix}.
\label{tmatrix_explicit}
\eeq 
In the presence of disorder there is no translational invariance and 
the transfer matrices $M_n$ will site dependent. The topological invariant can be constructed from the eigenvalues of the \textit{full} transfer matrix
\beq
M = \prod\limits_{n=1}^{N} M_n.
\eeq
In particular, one can show that the condition for the existence of one pair of Majorana modes at zero energy with 
normalizable wave function ($\sum_n |\psi_n|^2 <\infty$) corresponds to the existence of an odd number of 
eigenvalues of $M$ with magnitude less than\cite{DeGottardi11_MFs_with_disorder,DeGottardi_MFs_with_spatially_varying_potentials} 1. Equivalently, the number of roots of the characteristic polynomial $f(z)= \text{Det }(I- z M)$ lying inside the unit circle,
\beq
n_f = \frac{1}{2 \pi i}\int_{|z|=1} dz \frac{f'(z)}{f(z)},
\eeq
should be odd.
The above considerations give a concrete way to find the phase boundary between topological and non-topological regions in a closed system. 
In the clean case, where the transfer matrices $M_n$ are all equal, the physics of localization is determined by any of the $M_n$ matrices, and the well-known Pfaffian criterion for a topological phase transition in an isolated NW \cite{kitaev2001}, i.e., 
$\sqrt{(2t + \mu)^2 + \Delta^2} < V_z < \sqrt{(2t - \mu)^2 + \Delta^2}$, is recovered.

From the full transfer matrix we obtain the transmission matrix $tt^\dagger$  using the identity \cite{Beenakker97_Random_matrix_review}

\beq
[2 + M M^{\dagger} + (M M^{\dagger})^{-1}]^{-1}&=&\frac{1}{4}
\begin{pmatrix} 
t t^{\dagger} & 0 \\ 
0 & {t'}^{\dagger} t' 
\end{pmatrix}. 
\eeq 
The eigenvalues $T_n$  of the matrix $tt^\dagger$ are related to the Lyapunov coefficients as $T_n=1/\cosh^2 \lambda_n$. By taking the 
trace we then obtain the transmission probability across the NW $T_{1N} =\sum_n T_n \propto G_\text{th}$ as described in the main text.
\twocolumngrid

\bibliographystyle{apsrev} 
\def\urlprefix{}
\def\url#1{}

\bibliographystyle{apsrev}

\end{document}